\documentclass[pra,superscriptaddress,twocolumn]{revtex4}
\usepackage[dvips]{graphicx}

\begin{document}
\title{Magnetic trapping of buffer-gas cooled chromium atoms and
   prospects for the extension to paramagnetic molecules.}

\author{Joost M Bakker} \affiliation{Humboldt Universit\"at zu Berlin,
   Institut f{\"u}r Physik, Hausvogteiplatz 5-7,10117 Berlin, Germany}
\email{bakker@physik.hu-berlin.de} \affiliation{Fritz-Haber-Institut
   der Max-Planck-Gesellschaft, Faradayweg 4-6, D-14195 Berlin,
   Germany}

\author{Michael Stoll} \affiliation{Fritz-Haber-Institut der
   Max-Planck-Gesellschaft, Faradayweg 4-6, D-14195 Berlin, Germany}

\author{Dennis R Weise}\affiliation{Universit\"at Konstanz,
   Fachbereich Physik, 78457 Konstanz, Germany} 

\author{Oliver Vogelsang} \affiliation{Universit\"at Konstanz,
   Fachbereich Physik, 78457 Konstanz, Germany}

\author{Gerard Meijer}
\affiliation{Fritz-Haber-Institut der Max-Planck-Gesellschaft,
   Faradayweg 4-6, D-14195 Berlin, Germany} 

\author{Achim
   Peters}\affiliation{Humboldt Universit\"at zu Berlin, Institut
   f{\"u}r Physik, Hausvogteiplatz 5-7,10117 Berlin, Germany}







\date{\today}

\begin{abstract}
   
   We report the successful buffer-gas cooling and magnetic trapping
   of chromium atoms with densities exceeding $10^{12}$ atoms per
   cm$^{3}$ at a temperature of 350 mK for the trapped sample. The
   possibilities to extend the method to buffer-gas cool and
   magnetically trap molecules are discussed. To minimize the most
   important loss mechanism in magnetic trapping, molecules with a
   small spin-spin interaction and a large rotational constant are
   preferred. Both the CrH ($^6\Sigma^+$ ground state) and MnH
   ($^7\Sigma^+$) radicals appear to be suitable systems for future
   experiments.
   
\end{abstract}

\maketitle

\section{Introduction}
\label{sec:intro}
The buffer-gas cooling and magnetic trapping of CaH in 1998
constituted the first experiment where large samples of cold neutral
molecules were confined by externally applied electromagnetic fields
\cite{Weinstein:Nat395:148}. Since then, a number of alternative
methods has been developed to produce and capture cold molecules, such
as Stark deceleration of polar molecules in a molecular beam and
subsequent electric trapping \cite{Bethlem:Nat406:491},
photoassociation of laser-cooled atoms in a magneto-optical trap
\cite{Vanhaecke:PRL89:063001}, and molecule formation by tuning
through magnetic Feshbach resonances in an optical dipole trap
\cite{Jochim:PRL91:240402}. Among these approaches, the thermalization
with a cold buffer gas is the most general cooling method, as it is
applicable to virtually all atomic and molecular species. It also
offers the benefit of producing very large samples at sub-Kelvin
temperatures, which form a good starting point for further cooling by
other techniques, such as evaporative cooling.

Doyle and co-workers have successfully applied the buffer gas method
\cite{DeCarvalho:EPJD7:289} to trap atomic chromium and europium
\cite{Weinstein:PRA57:R3173,Kim:PRL78:3665} as well as several other
atoms \cite{Hancox:Nat431:281}. Apart from the above mentioned
experiments on CaH, they also demonstrated buffer-gas cooling of VO
and PbO \cite{Weinstein:JCP109:2656,Egorov:PRA63:030501} molecules.
Although highly successful, buffer gas cooling in combination with
magnetic trapping of neutral species up to now had only been
implemented by this one group.

Here we report on our buffer-gas loading experiment, aiming at the
production of large samples of trapped cold molecules. We have
validated the performance of the new system by buffer-gas cooling and
trapping of atomic chromium (Cr). Cr was chosen as it has a large
magnetic moment of 6 $\mu_B$ and strong, well-characterized electronic
transitions suitable for detection using absorption spectroscopy.
Moreover, the most abundant isotope $^{52}$Cr (86 \%) has zero nuclear
spin, which considerably simplifies the spectrum. As mentioned above,
Cr was also one of the first atoms to be magnetically trapped using
buffer-gas cooling by Doyle and co-workers, and it has been shown that
evaporative cooling to quantum degeneracy is feasible
\cite{Griesmaier:PRL94:160401}. In our current experiments we trap Cr
atoms and observe holding times exceeding one minute.

In the second part of the paper we describe the criteria we have used
to select the molecules to be studied with this new experimental
setup. We discuss the technological constraints as well as the
relevant intrinsic properties of molecules that determine the chances
of success of buffer-gas cooling experiments in conjunction with
magnetic trapping.

\section{Method and experimental setup}
\label{sec:expt}
Buffer-gas cooling of atoms and molecules is done by thermalizing them
with a cold background gas. Collisions with the background gas
particles take out small quantities of energy of the particles to be
cooled, and after many collisions the particles are thermalized with
the background gas. The choice of background gas is restricted to
those species that have at the (low) temperature of interest a large
enough density to facilitate a large number of collisions while not
reacting with the collision partner to form new species. Also, if
buffer gas cooling is combined with trapping using external fields,
the buffer gas should be insensitive to these fields. In practice, the
only gas that fulfills these conditions is helium.

At the start of experiments, a dense helium vapor is created in an
experimental cell. The buffer gas density is chosen such that hot
particles injected into it are cooled sufficiently to be trapped by an
inhomogeneous magnetic field before they can diffuse to the cell wall
and adsorb to it. On the other hand, the density must be low enough to
allow for the particles to reach the trap center. Optimal buffer gas
densities are typically in the range of 10$^{16}$-10$^{17}$ atoms per
cm$^{3}$, which corresponds to the vapor pressure of $^4$He at $\sim$
700 mK, or $^3$He at $\sim$ 300 mK \cite{DeCarvalho:EPJD7:289}.

After the creation of the required buffer gas density, the species to
be cooled is injected into the cell. This can be done through
capillary injection \cite{Messer:PRL53:2555}, or by ablation from a
solid state precursor material placed within that cell. In another
current project, efforts are made to use a molecular beam to inject
molecules into the buffer-gas \cite{Egorov:EJPD31:307}.  Upon
injection into the cell, the particles collide with the helium atoms
in the vapor. By thermalizing with the helium vapor, species in
low-field seeking (LFS) states are drawn into the local magnetic field
minimum; those in high-field seeking (HFS) states are repelled and
lost at the cell wall. To fully isolate the trapped sample, it is
necessary to remove the helium, as it forms a thermal link to the cell
wall. This can be done by cooling the cell down, thereby effectively
cryopumping the helium out.

For the cooled species to be magnetically confined, a (strong)
magnetic field around the experimental cell, with a field minimum
coinciding with the cell center, must be applied. This can be done
using a magnetic quadrupole trap consisting of two parallel coils
where counter-propagating currents create a zero magnetic field in the
center between the coils. Typically, the trap depths required for
buffer gas cooling should be on the order of a few Kelvin. As the trap
depth $E$ (in units of temperature) is given by $E=\mu H/k_{B}$, where
$\mu$ is the magnetic moment of the species $H$ the magnetic field
magnitude, and $k_{B}$ the Boltzmann constant, this leads to typically
required field strengths of $\sim$ 2 T. These fields can be generated
by coils of superconducting material, for which a cooling to liquid
helium temperatures is required.

\begin{figure}[t]
   \centering \includegraphics[width=1.0\columnwidth]{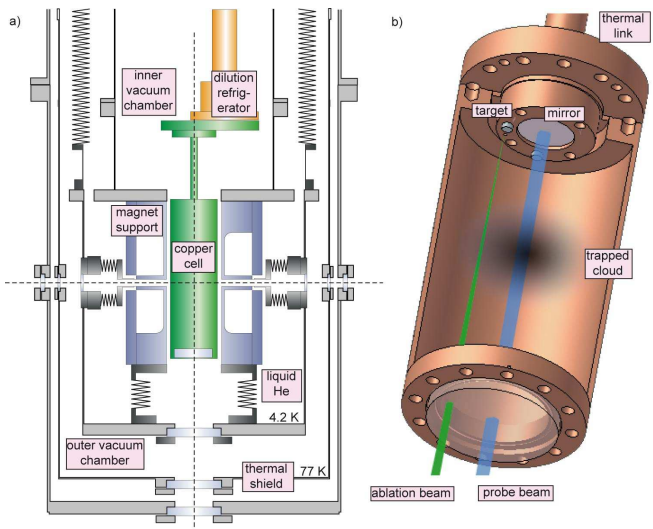}
\caption{Scheme of the experimental setup. Figure \ref{fig:setup}a
   shows the experimental cell within the custom-made cryostat,
   positioned within the magnet bore.  Figure \ref{fig:setup}b depicts
   a cut through the copper cell itself.}
\label{fig:setup}
\end{figure}

The refrigerator for cooling the experimental cell and the liquid
helium bath that holds the magnet are implemented in a custom-made
cryostat which is schematically depicted in Figure \ref{fig:setup}.
The cryostat is built around a commercial $^3$He-$^4$He dilution
refrigerator system (Leiden Cryogenics Minikelvin 126-700), with a
specified cooling power of 500 $\mu$W at 120 mK. The magnet consists
of two coils of NbTi/Cu wire spun onto a titanium bobbin that is
designed to withstand the repulsive forces of the two coils. The
magnet design is similar to the Mark-3 design by Doyle and co-workers
\cite{Harris:RSI75:17}, but allows for three axes along which the cell
can be optically interrogated: one vertical and two horizontal. From
the bottom of the cryostat, three apertures at the different
temperature stages of the cryostat (room temperature, 77 K and 4 K)
allow for the implementation of 75 mm diameter windows. From the top
of the cryostat and from the four lateral sides there are apertures
for 25 mm diameter windows. To be able to use the four side windows,
there are 10 mm diameter clearances through the magnet support
structure at the height of the trap center.
  
The experimental cell, made of oxygen-free high conductivity copper
(OFHC), is attached to the mixing chamber of the dilution refrigerator
and positioned in the center of the magnet bore. The cell is linked to
the cryostat by a 10 cm long copper rod of 15 mm diameter forming a
well-defined thermal impedance. Optical access into the cell is
provided by a 50 mm diameter window sealing the bottom of the cell. A
buffer gas fill line is implemented using a CuNi capillary. Finally,
the cell is equipped with RuO thermistors and a resistive heater for
temperature monitoring and control.

In the current experiment $^4$He is used as a buffer gas and the
appropriate density is created by bringing the entire cell to a
temperature of choice. The density is limited by the pre-defined
amount of helium let into the cell. Cr atoms are injected via
laser-ablation from a solid chip of Cr, positioned at the top of the
cell, using a $\sim$ 10 ns duration frequency-doubled Nd:YAG laser
pulse at 532 nm with a typical pulse energy of $\sim$ 15 mJ. We detect
Cr atoms in their $m_J=+3$ sublevel of the ground state by monitoring
absorption of light resonant with the $a^7S_3(m_J=+3) \leftarrow
z^7P_4(m_J=+4)$ transition around 425 nm. This is a nearly closed
transition where almost all excited Cr atoms decay into the initial
state (leakage $\sim 10^{-6}$). The light is generated by frequency
doubling the output of a grating-stabilized diode laser (Toptica
DL100) at 850 nm in a KNbO$_{3}$ crystal within an external bow-tie
cavity. The cavity is actively locked to the diode laser using a
Pound-Drever-Hall scheme \cite{Pound-Drever-Hall}. On long timescales
the cavity length is adjusted piezo-electrically to match the laser
frequency; on short time scales, the frequency of the diode laser is
adjusted to match the cavity length by modulating the diode current.
It must be stressed that the laser is not frequency-locked to an
external reference, so that long-term frequency drifts can occur. The
frequency-doubled light is amplitude stabilized by an acousto-optic
modulator (AOM), spatially filtered, expanded and (optionally) sent
through an aperture defining a beam diameter of a few mm.

This beam is then sent into the cell using a beam splitter and
retroreflected by a mirror mounted at the top of the cell. For the
current experiments only the windows that give access from the bottom
of the cell are implemented, and to minimize optical etalon effects
all windows are wedged by $\sim$ 1$^o$ and anti-reflection (AR) coated
for the wavelengths used in the experiment. Light exiting the cell is
directed onto a photomultiplier tube (PMT, Electron Tubes 9813QB). A
fraction of the light that is not entering the cryostat is directed
onto a second PMT (Electron Tubes 9814QB) for normalization and
correction of residual intensity fluctuations. Typical light powers
used are less than 1 $\mu$W, to avoid both heating of the cell and
loss of atoms due to optical pumping effects. The measured
transmission is thus obtained by normalizing the detection signal on
the reference signal. When the detection laser is scanned in
frequency, the signal is corrected for etalon effects by recording a
background transmission signal on which the signal is normalized. In
an alternative detection scheme, the transmitted beam is directly
steered onto the light-sensitive chip of a CCD camera (PCO Imaging,
Sensicam SuperVGA).

\section{Magnetic trapping of Chromium}
\label{sec:results}

\begin{figure}[t]
   \centering \includegraphics[width=1.0\columnwidth]{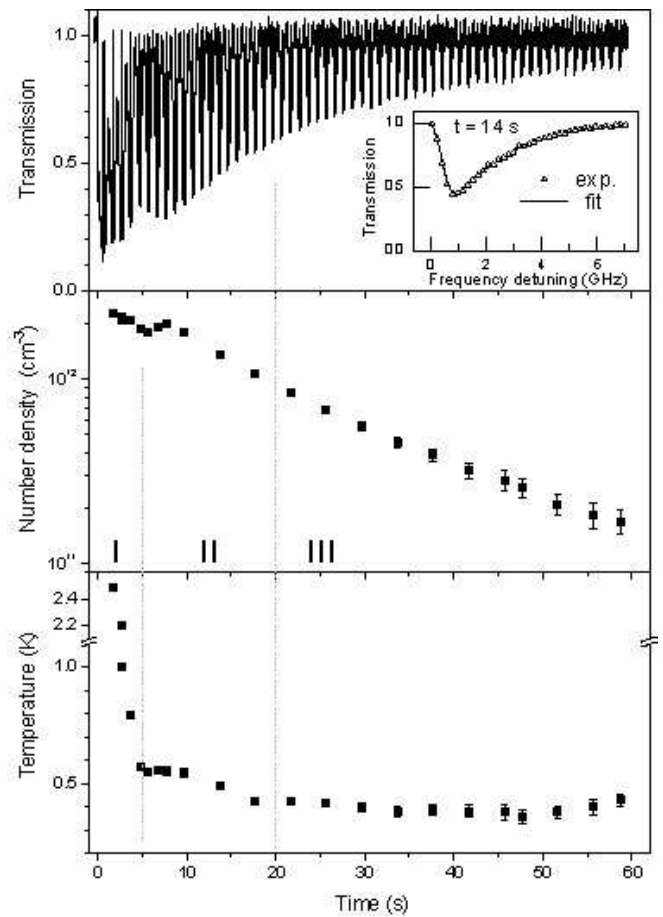}
   \caption{Transmission through a sample of Cr atoms in the buffer 
      gas cell (top graph). The probe laser frequency is scanned
      repeatedly over the atomic transition with a repetition rate of
      1 Hz. The inset shows a typical absorption profile (triangles)
      after $\sim$ 14 s, together with a simulation (dotted line). The
      other two graphs show the evolution of the particle density
      (middle) and temperature (bottom graph) of the trapped atom
      ensemble over time.}
   \label{fig:raw}
\end{figure}

Figure \ref{fig:raw}a depicts a typical recorded transmission trace
spanning a time period of 1 minute. In this experiment, starting at
t=-3.5 s, the cell is heated for a period of 3 s with a heating power
of $\sim$ 20 mW. Then, at t=0 s, the ablation laser ( 15 mJ) is fired
to inject Cr atoms into the helium vapor. The optimal loading
conditions are found in a procedure where the heating power and
ablation are optimized to yield the best capture efficiencies. A field
free measurement under the same conditions suggests that the heating
over a few seconds ensures a full evaporation of the helium off the
cell wall, whereas the short waiting period allows for the buffer gas
to be cooled somewhat without noticeable cryopumping effects taking
place. The 15 mJ used for the ablation is a compromise between signal
intensity and the temperature at which the atoms are captured in the
trap.

The trace shown in Figure \ref{fig:raw}a is recorded by repeatedly
scanning the probe laser frequency over a range of $\sim$ 10 GHz to
obtain a magnetic-field broadened absorption line while minimizing the
effect of long term drifts of the laser wavelength. In the scanned
signal, some interesting dynamics can be observed. In the initial few
seconds after ablation, the strong absorption signal first decreases,
then shortly grows again and finally decays slowly. Additionally, a
rapid narrowing of the line profile can be seen during the first 5
seconds following ablation. These different regimes will be discussed
in detail below.

More detailed information about the density and temperature evolution
of the trapped atom ensemble can be obtained by simulating the spectra
and comparing them to the recorded ones. First we note that at
temperatures below 1K the expected Doppler broadening is less than 100
MHz, which is negligible in comparison with the observed Zeeman
broadened linewidths of 1 GHz and larger. For the modeling, it is
assumed that the cloud of atoms is in thermal equilibrium and can be
characterized by a Boltzmann temperature \cite{Luiten:PRA53:381}. For
any given point in the trapping field, the Zeeman shifted transition
frequency and a (relative) population density and a resulting
absorption cross-section are calculated.  A transmission spectrum is
obtained by numerically integrating the absorption over the volume
defined by the probe beam. From the inset in Figure \ref{fig:raw}a it
can be seen that the simulated spectrum is in excellent agreement with
the recorded data.

The simulations show that the spectral shape strongly depends on the
probe beam parameters and we have thus carefully measured beam
diameter and intensity profile. However, the relative position of the
probe beam with respect to the center of the magnetic quadrupole trap
is less amenable to direct measurements and must therefore be included
in the model. This relative position is especially important when the
probe beam is linearly polarized, as the projection of the driving
$E$-field onto the quantization axis of the atoms, the inhomogeneous
magnetic field, has a strong angular dependence. It is rather
straightforward to derive that the angular dependence of the
absorption coefficient is $(\cos^2\theta + \sin^2\theta \sin^2\phi)$,
where $\theta$ is the angle between magnetic field and light
propagation axis, and $\phi$ that between polarization direction and
magnetic field, respectively. To the best of our knowledge, this full
angular dependence has not been implemented in modeling of data in
previous buffer-gas loading experiments. Best agreement is obtained
when it is assumed that our 2 mm diameter beam is offset by 0.3 mm
from the trap center, which is well within the limits of optical
alignment. The remaining free parameters used in the simulations are
the particle density at the trap center and the Boltzmann temperature,
which are found using a least-squares fitting procedure.
   
A possible error in the parameters obtained this way arises from a
breakdown of the assumption of thermal equilibrium. This is likely the
case at times shortly after the ablation pulse, where the changes in
temperature are on comparable time scales as the time it takes to
record one spectral profile. For this reason, the temperatures and
densities obtained before t=5 seconds are given without error bars and
should be considered as indicative only. Additionally, long-term
drifts in the laser frequency could amount to errors of a few \%. The
densities and temperatures of the sample as obtained from the spectral
simulations are depicted in the lower two graphs in Figure
\ref{fig:raw}. The size of the error bars reflects the reduced
signal-to-noise level at later times, when absorption signals become
less pronounced.

Data at short times after ablation have not been published by Doyle
and co-workers previously and it is interesting to speculate on the
nature of the observed dynamics. The time evolution can roughly be
divided into three periods which are indicated by I, II and III in
Figure \ref{fig:raw}b. During period I, the temperature falls rapidly
from $\sim$ 1.5 K at t=2 s to some 500 mK after 5 s. This is
attributed to so-called spilling effects, where during trap loading
the warmest particles are lost over the trap edge. As a result, the
overall temperature drops.

In Period II, the sample temperature does not change as dramatically
as in Period I. The sample appears to be in thermal equilibrium as it
has a temperature similar to that of the cell wall. The obvious {\it
   recurrence} in the transmission spectrum after 5 s is also observed
in the density. The increase in the density is interpreted as a net
diffusion of cooled atoms into the trap center.
 
After $\sim$ 20 s, the temperature of the cell wall (not shown) has
dropped to around 350 mK and the gas-phase helium within the cell is
now cryopumped to the cell wall. As a consequence the sample now
becomes thermally isolated and the temperature of the trapped cloud is
now no longer decreasing. Its final temperature thus corresponds to
the cell temperature at the time of thermal disconnect.

At times later than 20 s, a clear continuous loss of Cr is observed.
It is not expected that evaporation plays a large role at this point,
as the effective trap depth at these temperatures is very large and
evaporation rates are expected to decrease exponentially with
effective trap depth \cite{Ketterle:AAMOP37:181}. A further indication
for this is that the observed temperatures do not change, which rules
out effective evaporative cooling. As additionally the buffer gas
temperature is similar to or lower than the Cr temperature, we can
thus rule out elastic collisions as a loss mechanism.

As a consequence, the observed trap loss must be due to inelastic
collisions, which change the spin of the atom. Several such processes
are known and they can be distinguished by the number of Cr atoms
involved. The dominant loss mechanism may be found by fitting
functional forms to the number density evolution. In an earlier set of
data recorded over 180 s (where the signal was noisier and thus less
suitable for quantitative data analysis) a clear signature is found of
two-body loss processes, i.e. collisions in which two Cr atoms are
involved. For the data presented in Figure \ref{fig:raw}, the
signature of two body loss processes cannot be easily discerned from
that due to collisions with background gas. If we leave one-body decay
out of consideration, we obtain a maximum value for the two-body
inelastic collision cross-section of 6$\cdot$ 10$^{-13} cm^2$ at T=350
mK , consistent with a previous measurement
\cite{deCarvalho:JOSAB20:1131}.

\begin{figure}[t]
   \centering
   \includegraphics[width=1.0\columnwidth]{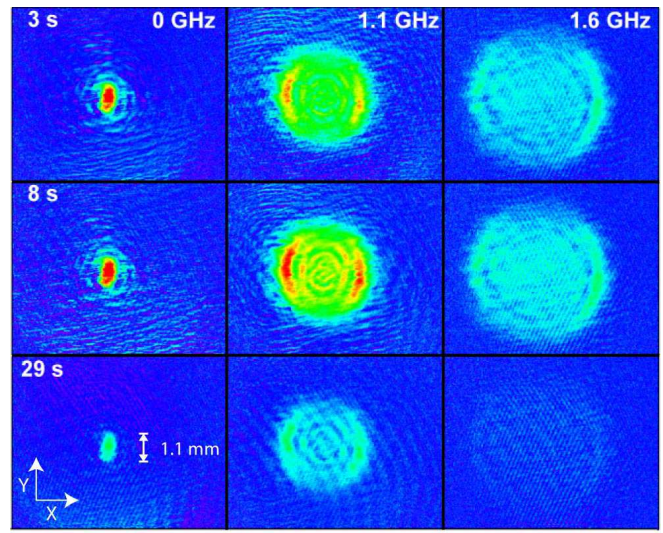}
   \caption{Absorption images of a trapped cloud of Cr atoms, at three 
      different times (columns) and at three different laser
      frequencies (rows). The size of the image is 8.6 mm x 6.9 mm. }
   \label{fig:ccd}
\end{figure}

In the measurements presented above, the probe laser diameter is
limited to $\sim$ 2 mm and is aligned close to the trap center. This
way not the whole trapping volume is addressed. In an alternative
experimental configuration, the full waist of the collimated probe
beam is sent through the trap center, retroreflected and steered
directly onto the light-sensitive element of a CCD camera. For these
imaging experiments the laser frequency is kept at a fixed value.
Since the atomic resonance frequency is determined by the
field-dependent Zeeman shift, a fixed frequency experiment selectively
probes for atoms in specific 'shells' within the trapping volume. The
results of these measurements are depicted in Figure \ref{fig:ccd} for
three probe laser frequencies and three different times at the maximum
trap depth of 6.8 Kelvin. The images shown (1 second gate time) are
obtained as the ratio of an image recorded with Cr in the trap and one
where no Cr is in the system. The size of the images is 8.6 mm
(horizontal axis) by 6.9 mm (vertical). Since the beam is collimated,
the spatial extent of the absorption signal on the CCD element has a
1:1 correspondence to the trapped cloud size.

The first set of images is recorded at the frequency of the field-free
atomic transition. A strong absorption signal is clearly visible at
the trap center.  One can see that the absorption is anisotropic,
which is a manifestation of the use of linearly polarized probe light.
When the probe laser is tuned towards higher frequencies, a ring
structure becomes apparent in the images. Here, it is obvious that the
angular intensity distribution is proportional to $\sin^2\phi$, as
discussed above. At a frequency-offset of +1.1 GHz, the strongest
absorption is observed in a ring with a radius of $\sim$ 1.5 mm. The
reason for such a pronounced ring is that the probe beam line-of-sight
here coincides with a 'shell' of Cr atoms trapped at the same field
magnitude. Again, the asymmetry attributed to the laser polarization
is clearly visible. The ring structure allows for a consistency check
of our trapping field calculations. If we assume that the probe beam
is collimated well, this implies that at a radius of 1.5 mm the
magnetic field calculated from the Zeeman shift $\Delta \nu=H\mu_B
(g_{f}m_{J,f}-g_{i}m_{J,i})/h$ is 76 mT. This is indeed consistent
with our trapping field calculations (69 mT at 1.5 mm), used in the
spectral simulations described earlier.

Apart from this strong outer ring of absorption, some weaker but still
pronounced rings are also visible. The origin of these rings is
unknown to us. It is unlikely that they are due to population in lower
$m_J$ states, as they persist for tens of seconds and it is expected
that population in lower $m_J$ states is lost more quickly than in the
$m_J=+3$ state due to a combination of lower trap depth and
spin-exchange collisions. It can be speculated that they are caused by
diffraction of the beam by the dense atomic cloud in the trap center.
It is, however, unfortunately not straightforward to simulate this
diffraction pattern due to our probe geometry where the probe beam
passes the trapped cloud twice. Without diffractive effects included,
a semi-quantitative agreement between our simulated data and the
images is found. We therefore do not extract temperature and density
information out of these images. For the absorption measurements where
only a small probe beam is steered through the center of the cloud,
such diffractive effects are not expected to play an important role as
the cloud size is larger than the beamsize, and the sensitive area of
the PMT is significantly larger than the probe beam diameter.

\begin{figure}[t]
   \centering
   \includegraphics[width=1.0\columnwidth]{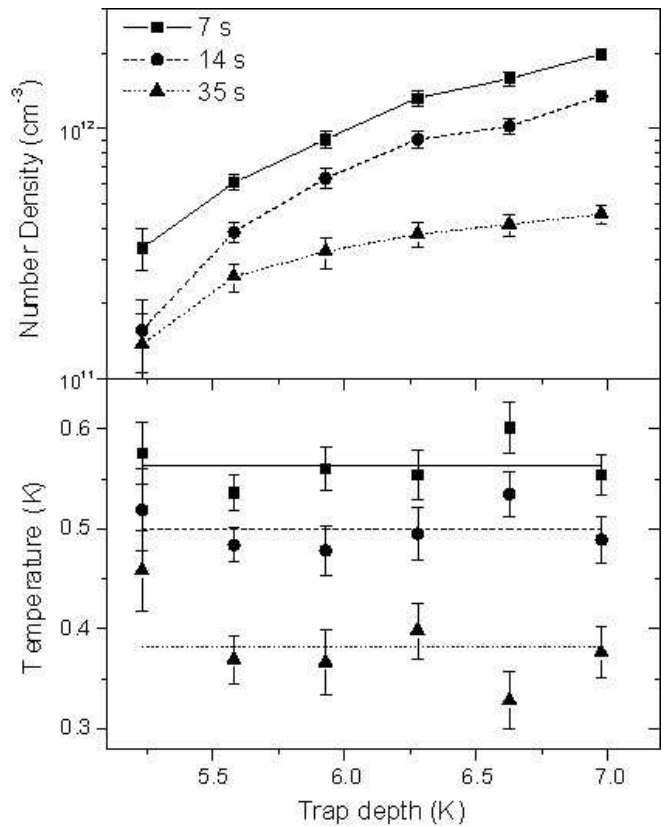}
\caption{Density (upper graph) and temperature (lower graph)
   of a trapped cloud of Cr atoms as a function of trap depth,
   measured at three different times; at t=7 s (squares), t=14 s
   (circles) and 35 s (triangles) after ablation. The lines in the
   upper graph are guides to the eye, whereas in the bottom graph they
   represent the average temperature.}
\label{fig:trapdepth}
\end{figure}

To investigate the potential of this system for future experiments on
the buffer gas cooling and trapping of molecular species, it is of
interest to evaluate the trapping efficiencies at lower trap depths.
Such a measurement is performed by lowering the current through the
two anti-Helmholtz coils. The results of these tests are depicted in
Figure \ref{fig:trapdepth} where the observed number densities and
temperatures of a trapped Cr cloud, evaluated at three different
times, are shown as a function of trap depth. Each value is obtained
by averaging over several measurements. The trends that can be seen
from these graphs are clear: a lower trap depth leads to a substantial
drop in the number of captured atoms. The temperature of the trapped
atoms is independent of trap depth. It seems thus obvious that a
shallower trap captures less particles, while the temperature behavior
gives us supporting evidence that no evaporative cooling, and thus no
evaporation takes place. It is obvious from this Figure that if a
molecule were to be trapped, it should possess a large enough magnetic
moment to allow for a trap depth of at least 5 K. Molecules that could
fulfill this condition are discussed in Section \ref{sec:molecule}.

\section{Search for a molecular candidate for buffer gas cooling}
\label{sec:molecule}

As He atoms are highly nonreactive and do not easily form new species,
we can consider the helium buffer gas cooling a universal cooling
method. As such, it has already been shown that one can use this
technique as a source of cold molecules \cite{Maxwell:PRL95:173201}.
However, if the cooling is to be combined with magnetic trapping of
the (paramagnetic) species, the universality is severely limited due
to several constraints. At the heart of these constraints lie
inelastic collisions, in which molecules in a low-field seeking (LFS)
state undergo a transition into a high-field seeking (HFS) state, a
process often referred to as Zeeman relaxation. The HFS molecules are
then rapidly ejected from the trapping region. The ratio of inelastic
to elastic collisions is thus crucial for the efficient transfer of
energy while maintaining the molecules in a trappable state. The
collision partners can be a molecule and a helium atom, but also two
molecules. The former case is crucial for the feasibility of
buffer-gas loading, while the latter governs a possible evaporative
cooling process to further lower the temperature and increase the
phase-space density of molecular samples.

Theoretical studies on the mechanisms of inelastic collisions with
helium atoms in external fields are few, and have been limited to
molecules in $^2\Sigma$ and $^3\Sigma$ ground states \cite{
   Volpi:PRA65:052712, Krems:PRA67:060703,Krems:PRA68:051401,
   Krems:JCP117:118,Cybulski:JCP122:094307}, although a study on
collisions between He and the OH radical (in a $^2\Pi$ state) is
forthcoming \cite{Groenenboom:privcom}. For $^2\Sigma$ molecules the
Zeeman relaxation occurs through spin-rotational coupling between the
rotational ground state and rotationally excited states
\cite{Krems:JCP120:2296}. Calculations for CaF and CaH molecules have
shown that Zeeman relaxation rates are strongly dependent on the
spin-rotation interaction constant $\gamma_{sr}$ and decrease with the
energy difference between rotational states. The experimental data
available for these molecules confirm this
\cite{Weinstein:Nat395:148,Maussang:PRL94:123002}.

For higher-spin states, no experimental data is available. It is,
however, predicted that the Zeeman relaxation rates are considerably
higher. In $^3\Sigma$ states, the spin-spin interaction is the
dominant factor and again a larger energy difference between the
rotational states decreases the couplings between LFS and HFS states.
As a consequence, the spin-spin interaction parameter $\lambda_{ss}$
should be relatively small and the rotational constant large to
successfully trap molecules \cite{Krems:JCP120:2296}. For the NH
($^3\Sigma^-$) radical, the calculated ratio of elastic to inelastic
collisions has been shown to be favorable for trapping experiments
\cite{Cybulski:JCP122:094307}.

About the collisional properties of molecules in ground states other
than $^2\Sigma$ or $^3\Sigma$ little is known. An unsuccessful attempt
was made to buffer gas cool and trap VO molecules ($^4\Sigma^-$ ground
state), but it is unclear whether Zeeman relaxation was a problem
\cite{Weinstein:JCP109:2656}. Theoretical calculations that predict
collisional properties of such high-spin molecules are unknown to us.
Based on the lack of such data, we here propose to buffer gas cool
molecules with a high-spin ground state to investigate their
suitability for magnetic trapping and possible evaporative cooling
experiments. By extrapolating the cases built for $^2\Sigma$ and
$^3\Sigma$ molecules, we assume that the candidate molecule should
possess a large rotational constant B and a small spin-spin
interaction constant $\lambda_{ss}$ to minimize spin-spin couplings.
As diatomics containing hydrogen atoms have the largest rotational
constants, metal hydrides are a natural choice.

Apart from such general criteria more practical considerations also
play a role. As is observed above, the trap depth plays a crucial role
in the possibility to confine species over longer time periods. In the
present experiment, trap depths of 5.3 K and above have been employed
with success. A useful quantity to describe trapping is the capture
efficiency parameter $\eta$, defined as the ratio of trap depth to
sample temperature; $\eta = \mu H_{max}/k_{B}T_{mol} $, where $\mu$ is the
molecular magnetic moment, $H_{max}$ the magnetic field strength at
the trap edge and $T_{mol}$ the Boltzmann temperature of the molecular
sample. 
In the current experiment the trap edge is formed by the cell wall, 
where the highest achievable magnetic field is 1.7 T.
Thus, at the observed initial temperatures of 1.4 K (evaluated
at t=2 s), $\eta$ should be larger than 3.8 for a successful (initial)
confinement of a substantial sample of molecules.  This yields a minimum
magnetic moment of about 4.5 $\mu_B$. 

\begin{table}[tb]
\centering 

\caption{Diatomic molecules with high-spin electronic ground states
   and their rotational constant (B), spin-spin interaction constant
   ($\lambda_{ss}$) and spin-rotation interaction constant ($\gamma_{sr}$). All
   numbers are given in cm$^{-1}$. The ratio of $\lambda_{ss}$ over B is an
   indicator for the strength of Zeeman relaxation.}
\begin{tabular}{lc|rrlrr}
\hline
& & $B$ & $\lambda_{ss}$ & $|\lambda_{ss}/B|$&$\gamma_{sr}$&ref.\\
& &&&&$\times 10^3$&\\
\hline
\hline
GdO &$^9\Sigma^-   $& 0.355& -0.104& 0.29 &    0.1&\cite{Kaledin:JMS165:323}\\
MnH &$^7\Sigma^+ $& 5.606& -0.004& 7.2 10$^{-4}$ &   31.3&\cite{Gordon:JMS229:145}\\
MnF &$^7\Sigma^+ $& 0.353& -0.005& 1.3 10$^{-2}$ & 0.6&\cite{Sheridan:CPL380:632}\\
MnCl&$^7\Sigma^+ $& 0.158&  0.037& 0.24 &    0.4&\cite{Halfen:JCP122:054309}\\
CrH &$^6\Sigma^+ $& 6.132&  0.233& 3.8 10$^{-2}$ &  50.3&\cite{Corkery:JMS149:257}\\
MnO &$^6\Sigma^+ $& 0.501&  0.574& 1.1 &  -2.4&\cite{Namiki:JCP107:8848}\\
CrF &$^6\Sigma^+ $& 0.379&  0.539& 1.4 & 13.6 &\cite{Katoh:JCP120:7927}\\
MnS &$^6\Sigma^+ $& 0.195&  0.350& 1.8 & -2.4 &\cite{Thompsen:JCP116:10212}\\
CrCl&$^6\Sigma^+ $& 0.167&  0.266& 1.6 &   2.2&\cite{Katoh:JCP120:7927}\\
CrO &$^5\Pi_r    $& 0.524&  1.148& 2.2 &  10.5&\cite{Sheridan:ApJ576:1108}\\
CrN &$^4\Sigma^- $& 0.624&  2.611& 4.2 &   7.0&\cite{Sheridan:ApJ576:1108}\\
VO  &$^4\Sigma^- $& 0.546&  2.031& 3.7 &  22.5&\cite{Cheung:JMS91:165}\\
NbO &$^4\Sigma^- $& 0.432&  15.58& 36  &  334.0 &\cite{Adam:JCP100:6240}\\
TiH &$^4\Phi     $& 5.362&       & & 182.3&\cite{Launila:JCP104:6418}\\
\hline
NH  &$^3\Sigma^-   $&16.343&  0.920& 5.6  10$^{-2}$ &  54.7& \cite{Brazier:jms120:381}\\
O$_2$&$^3\Sigma^-  $& 1.438&  1.985& 1.4 & -8.4&\cite{Berden:PRA58:3114}\\
CaH &$^2\Sigma^+  $& 4.229&  &  & 46.7& \cite{Barclay:ApJ408:L65}\\
CaF &$^2\Sigma^+  $& 0.339&  &  &1.3&\cite{Maussang:PRL94:123002}\\
\hline 
\end{tabular}
\label{tab:molecules}
\end{table}

Based on the above mentioned criteria, we have performed an extensive
literature search to find an interesting and suitable molecular
candidate for buffer gas cooling and magnetic trapping. The results
are collected in Table \ref{tab:molecules}. To compile this table,
only molecules have been listed that possess an electronic ground
state with at least a fourfold spin multiplicity. For comparison, the
molecules studied so far using buffer-gas cooling have been included:
CaH, NH, CaF as well as the oxygen molecule ($^{16}$O$^{18}$O), a long
time prime candidate.  As a further restriction, only molecules that
have been observed in experimental work and for which an experimental
value of the rotational constant is available are included.  A column
is added with the ratio of the spin-spin interaction and rotational
constants, $|\lambda_{ss}/B|$, to guide the selection process in
minimizing the expected effects of Zeeman relaxation.

An inspection of this table indicates that there are three molecules
that clearly stand out in terms of expected stability against Zeeman
relaxation: CrH in its $^6\Sigma^+$ ground state, MnH ($^7\Sigma^+$)
and MnF ($^7\Sigma^+$). All three molecules possess a large magnetic
moment, which is nominally 5 and 6 $\mu_B$, respectively. We
disregard the MnF molecule, as it has a substantially smaller
rotational constant, which would result in level crossings between
different rotational manifolds at fields lower than 0.25 T
\footnote{The magnetic field at which the first rotational level
   crossings can occur can be approximated by $H_{cross}=2B/\mu$}. One
could point out that the TiH molecule also possesses a large magnetic
moment and rotational constant that should avoid these level
crossings; however, it is believed that non-$\Sigma$ state molecules
are highly susceptible to Zeeman relaxation, as direct couplings
between Zeeman states tend to become large \cite{Groenenboom:privcom}.
Since in addition no value of the spin-spin interaction constant could
be found for TiH, it is left out of consideration.

Both CrH and MnH fulfill the requirements of a large rotational
constant, a small spin-spin interaction constant while at the same
time having a high-spin ground state which should be amenable to
magnetic trapping. Both CrH and MnH were first observed in the 1930`s
by Gaydon and Pearse
\cite{Gaydon:Nat140:110,Pearse:ProcPhysSoc50:201}. CrH has by far
attracted the most experimental attention as a prototype of a
high-spin diatomic molecule, especially since it was discovered in the
emission spectra of L-type brown dwarfs \cite{Kirkpatrick:APJ519:802};
for an overview, see \cite{Shin:ApJ619:407} and references therein.
Because of its high-spin ground state, CrH is employed as a probe for
magnetic field strengths in sunspots \cite{Engvold:AAsuplser42:209}.
The magnetic parameters of CrH are well-established in the
X$^6\Sigma^+$ electronic ground state by FIR laser-magnetic resonance
spectroscopy \cite{Corkery:JMS149:257}. Recently, we also measured the
Zeeman spectrum of the A$^6\Sigma^+ \leftarrow$X$^6\Sigma^+$ (1-0)
transition around 767 nm, thereby determining the spectroscopic
parameters of the excited state that are needed to simulate spectra
recorded in an inhomogeneous magnetic field \cite{Chen-inprep}. For
MnH, the experimental studies are more sparse, which is likely due to
the highly complex structure of the spectroscopic transitions which
are split into numerous hyperfine components. The A$^7\Pi^+
\leftarrow$X$^7\Sigma^+$ (0-0) and (1-0) transitions near 624 and 568
nm, respectively, are well characterized
\cite{Varberg:jcp92:7123,Varberg:JCP95:1563,Varberg:JMS156:296} and
the ground state has been studied using ESR spectroscopy
\cite{Vanzee:JCP84:5968,Vanzee:JCP85:3237}.

Based on the extensive experimental information available, it seems
that the CrH radical is currently the most promising high-spin
candidate for studying the mechanisms involved in buffer gas cooling,
although it can not be ruled out that the significantly lower
spin-spin interaction of MnH will, in the end, favour this molecule.

\section{Conclusions}
\label{sec:conc}
We have demonstrated buffer-gas loading of paramagnetic species into a
magnetic trap. In first experiments we confined Cr atoms with storage
times of more than one minute. The use of an absorption based
detection scheme combined with a simulation of the absorption
properties of the cloud in the presence of the inhomogeneous magnetic
trapping field allows us to extract relevant parameters. Given storage
times of more than one minute, thermal contact between the trapped
cloud and the walls of the experimental cell is reliably broken after
a period of 20 s. The final temperatures of $\sim$ 350 mK reached are
determined by the temperature of the buffer gas at thermal disconnect.
Observed initial number densities exceed 10$^{12}$ per cm$^3$. We have
also succeeded in implementing continuous non-destructive absorption
imaging of the trapped cloud, which can provide complementary
information on temperature and internal state distribution over the
whole trapping volume.

Progressing towards the preparation of cold molecules, we have
outlined the criteria for selecting a molecular candidate to study in
buffer gas cooling experiments. Such a molecule should have, apart
from a high magnetic moment, a low cross-section for spin-changing
collisions which would result in trap loss. Due to their large
rotational constants and small spin-spin interaction coefficients, the
MnH and CrH molecules have been identified as potentially suitable
species. We will thus proceed by experimentally investigating the
relevant collisional properties of these molecules. Measurements of
the effect of trap depth on the capture efficiency show that, if the
spin-changing collision rates are sufficiently low, the current
experimental setup should allow for trapping and thermally isolating
CrH and MnH.

{\bf Acknowledgment}

We gratefully acknowledge the support by the Deutsche Forschungs
Gemeinschaft (Forschergruppe Quantengase FOR282/2-1TP8). JMB
acknowledges funding through the European Network for Cold Molecules
(RTN2-2001-00498). We thank Bretislav Friedrich and John M. Doyle for
stimulating discussions and their help in constructing this
experiment. Last but not least we want to thank Tim Steimle for
bringing MnH to our attention, and the referee for helpful comments
that considerably improved the manuscript.

\vspace{1cm}

\bibliographystyle{apsrev}
\bibliography{string,mp,atoms,XH-radicals,buffergas,bas,general,misc}

\begin{thebibliography}{53}
\expandafter\ifx\csname natexlab\endcsname\relax\def\natexlab#1{#1}\fi
\expandafter\ifx\csname bibnamefont\endcsname\relax
  \def\bibnamefont#1{#1}\fi
\expandafter\ifx\csname bibfnamefont\endcsname\relax
  \def\bibfnamefont#1{#1}\fi
\expandafter\ifx\csname citenamefont\endcsname\relax
  \def\citenamefont#1{#1}\fi
\expandafter\ifx\csname url\endcsname\relax
  \def\url#1{\texttt{#1}}\fi
\expandafter\ifx\csname urlprefix\endcsname\relax\def\urlprefix{URL }\fi
\providecommand{\bibinfo}[2]{#2}
\providecommand{\eprint}[2][]{\url{#2}}

\bibitem[{\citenamefont{Weinstein
  et~al.}(1998{\natexlab{a}})\citenamefont{Weinstein, {d}e{C}arvalho, Guillet,
  Friedrich, and Doyle}}]{Weinstein:Nat395:148}
\bibinfo{author}{\bibfnamefont{J.~D.} \bibnamefont{Weinstein}},
  \bibinfo{author}{\bibfnamefont{R.}~\bibnamefont{{d}e{C}arvalho}},
  \bibinfo{author}{\bibfnamefont{T.}~\bibnamefont{Guillet}},
  \bibinfo{author}{\bibfnamefont{B.}~\bibnamefont{Friedrich}},
  \bibnamefont{and} \bibinfo{author}{\bibfnamefont{J.~M.} \bibnamefont{Doyle}},
  \bibinfo{journal}{Nature} \textbf{\bibinfo{volume}{395}},
  \bibinfo{pages}{148} (\bibinfo{year}{1998}{\natexlab{a}}).

\bibitem[{\citenamefont{Bethlem et~al.}(2000)\citenamefont{Bethlem, Berden,
  Crompvoets, Jongma, van Roij, and Meijer}}]{Bethlem:Nat406:491}
\bibinfo{author}{\bibfnamefont{H.~L.} \bibnamefont{Bethlem}},
  \bibinfo{author}{\bibfnamefont{G.}~\bibnamefont{Berden}},
  \bibinfo{author}{\bibfnamefont{F.~M.~H.} \bibnamefont{Crompvoets}},
  \bibinfo{author}{\bibfnamefont{R.~T.} \bibnamefont{Jongma}},
  \bibinfo{author}{\bibfnamefont{A.~J.~A.} \bibnamefont{van Roij}},
  \bibnamefont{and} \bibinfo{author}{\bibfnamefont{G.}~\bibnamefont{Meijer}},
  \bibinfo{journal}{Nature} \textbf{\bibinfo{volume}{406}},
  \bibinfo{pages}{491} (\bibinfo{year}{2000}).

\bibitem[{\citenamefont{Vanhaecke et~al.}(2002)\citenamefont{Vanhaecke,
  de~Souza~Melo, Tolra, Comparat, and Pillet}}]{Vanhaecke:PRL89:063001}
\bibinfo{author}{\bibfnamefont{N.}~\bibnamefont{Vanhaecke}},
  \bibinfo{author}{\bibfnamefont{W.}~\bibnamefont{de~Souza~Melo}},
  \bibinfo{author}{\bibfnamefont{B.~L.} \bibnamefont{Tolra}},
  \bibinfo{author}{\bibfnamefont{D.}~\bibnamefont{Comparat}}, \bibnamefont{and}
  \bibinfo{author}{\bibfnamefont{P.}~\bibnamefont{Pillet}},
  \bibinfo{journal}{Phys. Rev. Lett.} \textbf{\bibinfo{volume}{89}},
  \bibinfo{pages}{063001} (\bibinfo{year}{2002}).

\bibitem[{\citenamefont{Jochim et~al.}(2003)\citenamefont{Jochim, Bartenstein,
  Altmeyer, Hendl, Chin, Denschlag, and Grimm}}]{Jochim:PRL91:240402}
\bibinfo{author}{\bibfnamefont{S.}~\bibnamefont{Jochim}},
  \bibinfo{author}{\bibfnamefont{M.}~\bibnamefont{Bartenstein}},
  \bibinfo{author}{\bibfnamefont{A.}~\bibnamefont{Altmeyer}},
  \bibinfo{author}{\bibfnamefont{G.}~\bibnamefont{Hendl}},
  \bibinfo{author}{\bibfnamefont{C.}~\bibnamefont{Chin}},
  \bibinfo{author}{\bibfnamefont{J.~H.} \bibnamefont{Denschlag}},
  \bibnamefont{and} \bibinfo{author}{\bibfnamefont{R.}~\bibnamefont{Grimm}},
  \bibinfo{journal}{Phys. Rev. Lett.} \textbf{\bibinfo{volume}{91}},
  \bibinfo{pages}{240402} (\bibinfo{year}{2003}).

\bibitem[{\citenamefont{deCarvalho et~al.}(1999)\citenamefont{deCarvalho,
  Doyle, Friedrich, Guillet, Kim, Patterson, and
  Weinstein}}]{DeCarvalho:EPJD7:289}
\bibinfo{author}{\bibfnamefont{R.}~\bibnamefont{deCarvalho}},
  \bibinfo{author}{\bibfnamefont{J.~M.} \bibnamefont{Doyle}},
  \bibinfo{author}{\bibfnamefont{B.}~\bibnamefont{Friedrich}},
  \bibinfo{author}{\bibfnamefont{T.}~\bibnamefont{Guillet}},
  \bibinfo{author}{\bibfnamefont{J.}~\bibnamefont{Kim}},
  \bibinfo{author}{\bibfnamefont{D.}~\bibnamefont{Patterson}},
  \bibnamefont{and} \bibinfo{author}{\bibfnamefont{J.~D.}
  \bibnamefont{Weinstein}}, \bibinfo{journal}{Eur. Phys. J. D}
  \textbf{\bibinfo{volume}{7}}, \bibinfo{pages}{289} (\bibinfo{year}{1999}).

\bibitem[{\citenamefont{Weinstein
  et~al.}(1998{\natexlab{b}})\citenamefont{Weinstein, deCarvalho, Kim,
  Patterson, Friedrich, and Doyle}}]{Weinstein:PRA57:R3173}
\bibinfo{author}{\bibfnamefont{J.~D.} \bibnamefont{Weinstein}},
  \bibinfo{author}{\bibfnamefont{R.}~\bibnamefont{deCarvalho}},
  \bibinfo{author}{\bibfnamefont{J.}~\bibnamefont{Kim}},
  \bibinfo{author}{\bibfnamefont{D.}~\bibnamefont{Patterson}},
  \bibinfo{author}{\bibfnamefont{B.}~\bibnamefont{Friedrich}},
  \bibnamefont{and} \bibinfo{author}{\bibfnamefont{J.~M.} \bibnamefont{Doyle}},
  \bibinfo{journal}{Phys. Rev. A} \textbf{\bibinfo{volume}{57}},
  \bibinfo{pages}{R3173} (\bibinfo{year}{1998}{\natexlab{b}}).

\bibitem[{\citenamefont{Kim et~al.}(1997)\citenamefont{Kim, Friedrich, Katz,
  Patterson, Weinstein, deCarvalho, and Doyle}}]{Kim:PRL78:3665}
\bibinfo{author}{\bibfnamefont{J.}~\bibnamefont{Kim}},
  \bibinfo{author}{\bibfnamefont{B.}~\bibnamefont{Friedrich}},
  \bibinfo{author}{\bibfnamefont{D.~P.} \bibnamefont{Katz}},
  \bibinfo{author}{\bibfnamefont{D.}~\bibnamefont{Patterson}},
  \bibinfo{author}{\bibfnamefont{J.~D.} \bibnamefont{Weinstein}},
  \bibinfo{author}{\bibfnamefont{R.}~\bibnamefont{deCarvalho}},
  \bibnamefont{and} \bibinfo{author}{\bibfnamefont{J.~M.} \bibnamefont{Doyle}},
  \bibinfo{journal}{Phys. Rev. Lett.} \textbf{\bibinfo{volume}{78}},
  \bibinfo{pages}{3665} (\bibinfo{year}{1997}).

\bibitem[{\citenamefont{Hancox et~al.}(2004)\citenamefont{Hancox, Doret,
  Hummon, Luo, and Doyle}}]{Hancox:Nat431:281}
\bibinfo{author}{\bibfnamefont{C.~I.} \bibnamefont{Hancox}},
  \bibinfo{author}{\bibfnamefont{S.~C.} \bibnamefont{Doret}},
  \bibinfo{author}{\bibfnamefont{M.~T.} \bibnamefont{Hummon}},
  \bibinfo{author}{\bibfnamefont{L.~J.} \bibnamefont{Luo}}, \bibnamefont{and}
  \bibinfo{author}{\bibfnamefont{J.~M.} \bibnamefont{Doyle}},
  \bibinfo{journal}{Nature} \textbf{\bibinfo{volume}{431}},
  \bibinfo{pages}{281} (\bibinfo{year}{2004}).

\bibitem[{\citenamefont{Weinstein
  et~al.}(1998{\natexlab{c}})\citenamefont{Weinstein, deCarvalho, Amar, Boca,
  Odom, Friedrich, and Doyle}}]{Weinstein:JCP109:2656}
\bibinfo{author}{\bibfnamefont{J.~D.} \bibnamefont{Weinstein}},
  \bibinfo{author}{\bibfnamefont{R.}~\bibnamefont{deCarvalho}},
  \bibinfo{author}{\bibfnamefont{K.}~\bibnamefont{Amar}},
  \bibinfo{author}{\bibfnamefont{A.}~\bibnamefont{Boca}},
  \bibinfo{author}{\bibfnamefont{B.~C.} \bibnamefont{Odom}},
  \bibinfo{author}{\bibfnamefont{B.}~\bibnamefont{Friedrich}},
  \bibnamefont{and} \bibinfo{author}{\bibfnamefont{J.~M.} \bibnamefont{Doyle}},
  \bibinfo{journal}{J. Chem. Phys.} \textbf{\bibinfo{volume}{109}},
  \bibinfo{pages}{2656} (\bibinfo{year}{1998}{\natexlab{c}}).

\bibitem[{\citenamefont{Egorov et~al.}(2001)\citenamefont{Egorov, Weinstein,
  Patterson, Friedrich, and Doyle}}]{Egorov:PRA63:030501}
\bibinfo{author}{\bibfnamefont{D.}~\bibnamefont{Egorov}},
  \bibinfo{author}{\bibfnamefont{J.~D.} \bibnamefont{Weinstein}},
  \bibinfo{author}{\bibfnamefont{D.}~\bibnamefont{Patterson}},
  \bibinfo{author}{\bibfnamefont{B.}~\bibnamefont{Friedrich}},
  \bibnamefont{and} \bibinfo{author}{\bibfnamefont{J.~M.} \bibnamefont{Doyle}},
  \bibinfo{journal}{Phys. Rev. A} \textbf{\bibinfo{volume}{63}},
  \bibinfo{pages}{030501} (\bibinfo{year}{2001}).

\bibitem[{\citenamefont{Griesmaier et~al.}(2005)\citenamefont{Griesmaier,
  Werner, Hensler, Stuhler, and Pfau}}]{Griesmaier:PRL94:160401}
\bibinfo{author}{\bibfnamefont{A.}~\bibnamefont{Griesmaier}},
  \bibinfo{author}{\bibfnamefont{J.}~\bibnamefont{Werner}},
  \bibinfo{author}{\bibfnamefont{S.}~\bibnamefont{Hensler}},
  \bibinfo{author}{\bibfnamefont{J.}~\bibnamefont{Stuhler}}, \bibnamefont{and}
  \bibinfo{author}{\bibfnamefont{T.}~\bibnamefont{Pfau}},
  \bibinfo{journal}{Phys. Rev. Lett.} \textbf{\bibinfo{volume}{94}},
  \bibinfo{pages}{160401} (\bibinfo{year}{2005}).

\bibitem[{\citenamefont{Messer and Delucia}(1984)}]{Messer:PRL53:2555}
\bibinfo{author}{\bibfnamefont{J.~K.} \bibnamefont{Messer}} \bibnamefont{and}
  \bibinfo{author}{\bibfnamefont{F.~C.} \bibnamefont{Delucia}},
  \bibinfo{journal}{Phys. Rev. Lett.} \textbf{\bibinfo{volume}{53}},
  \bibinfo{pages}{2555} (\bibinfo{year}{1984}).

\bibitem[{\citenamefont{Egorov et~al.}(2004)\citenamefont{Egorov, Campbell,
  Friedrich, Maxwell, Tsikata, van Buuren, and Doyle}}]{Egorov:EJPD31:307}
\bibinfo{author}{\bibfnamefont{D.}~\bibnamefont{Egorov}},
  \bibinfo{author}{\bibfnamefont{W.~C.} \bibnamefont{Campbell}},
  \bibinfo{author}{\bibfnamefont{B.}~\bibnamefont{Friedrich}},
  \bibinfo{author}{\bibfnamefont{S.~E.} \bibnamefont{Maxwell}},
  \bibinfo{author}{\bibfnamefont{E.}~\bibnamefont{Tsikata}},
  \bibinfo{author}{\bibfnamefont{L.~D.} \bibnamefont{van Buuren}},
  \bibnamefont{and} \bibinfo{author}{\bibfnamefont{J.~M.} \bibnamefont{Doyle}},
  \bibinfo{journal}{Eur. Phys. J. D} \textbf{\bibinfo{volume}{31}},
  \bibinfo{pages}{307} (\bibinfo{year}{2004}).

\bibitem[{\citenamefont{Harris et~al.}(2004)\citenamefont{Harris, Michniak,
  Nguyen, Campbell, Egorov, Maxwell, van Buuren, and Doyle}}]{Harris:RSI75:17}
\bibinfo{author}{\bibfnamefont{J.~G.~E.} \bibnamefont{Harris}},
  \bibinfo{author}{\bibfnamefont{R.~A.} \bibnamefont{Michniak}},
  \bibinfo{author}{\bibfnamefont{S.~V.} \bibnamefont{Nguyen}},
  \bibinfo{author}{\bibfnamefont{W.~C.} \bibnamefont{Campbell}},
  \bibinfo{author}{\bibfnamefont{D.}~\bibnamefont{Egorov}},
  \bibinfo{author}{\bibfnamefont{S.~E.} \bibnamefont{Maxwell}},
  \bibinfo{author}{\bibfnamefont{L.~D.} \bibnamefont{van Buuren}},
  \bibnamefont{and} \bibinfo{author}{\bibfnamefont{J.~M.} \bibnamefont{Doyle}},
  \bibinfo{journal}{Rev. Sci. Instrum.} \textbf{\bibinfo{volume}{75}},
  \bibinfo{pages}{17} (\bibinfo{year}{2004}).

\bibitem[{\citenamefont{Drever et~al.}(1983)\citenamefont{Drever, Hall,
  Kowalski, Hough, Ford, Munley, and Ward}}]{Pound-Drever-Hall}
\bibinfo{author}{\bibfnamefont{R.~W.~P.} \bibnamefont{Drever}},
  \bibinfo{author}{\bibfnamefont{J.~L.} \bibnamefont{Hall}},
  \bibinfo{author}{\bibfnamefont{F.~V.} \bibnamefont{Kowalski}},
  \bibinfo{author}{\bibfnamefont{J.}~\bibnamefont{Hough}},
  \bibinfo{author}{\bibfnamefont{G.~M.} \bibnamefont{Ford}},
  \bibinfo{author}{\bibfnamefont{A.~J.} \bibnamefont{Munley}},
  \bibnamefont{and} \bibinfo{author}{\bibfnamefont{H.}~\bibnamefont{Ward}},
  \bibinfo{journal}{Appl. Phys. B} \textbf{\bibinfo{volume}{31}},
  \bibinfo{pages}{97} (\bibinfo{year}{1983}).

\bibitem[{\citenamefont{Luiten et~al.}(1996)\citenamefont{Luiten, Reynolds, and
  Walraven}}]{Luiten:PRA53:381}
\bibinfo{author}{\bibfnamefont{O.~J.} \bibnamefont{Luiten}},
  \bibinfo{author}{\bibfnamefont{M.~W.} \bibnamefont{Reynolds}},
  \bibnamefont{and} \bibinfo{author}{\bibfnamefont{J.~T.~M.}
  \bibnamefont{Walraven}}, \bibinfo{journal}{Phys. Rev. A}
  \textbf{\bibinfo{volume}{53}}, \bibinfo{pages}{381} (\bibinfo{year}{1996}).

\bibitem[{\citenamefont{Ketterle and van Druten}(1996)}]{Ketterle:AAMOP37:181}
\bibinfo{author}{\bibfnamefont{W.}~\bibnamefont{Ketterle}} \bibnamefont{and}
  \bibinfo{author}{\bibfnamefont{N.~J.} \bibnamefont{van Druten}},
  \bibinfo{journal}{Adv. Atom. Mol. Opt. Phys.} \textbf{\bibinfo{volume}{37}},
  \bibinfo{pages}{181} (\bibinfo{year}{1996}).

\bibitem[{\citenamefont{deCarvalho et~al.}(2003)\citenamefont{deCarvalho,
  Hancox, and Doyle}}]{deCarvalho:JOSAB20:1131}
\bibinfo{author}{\bibfnamefont{R.}~\bibnamefont{deCarvalho}},
  \bibinfo{author}{\bibfnamefont{C.~I.} \bibnamefont{Hancox}},
  \bibnamefont{and} \bibinfo{author}{\bibfnamefont{J.~M.} \bibnamefont{Doyle}},
  \bibinfo{journal}{J. Opt. Soc. Am. B} \textbf{\bibinfo{volume}{20}},
  \bibinfo{pages}{1131} (\bibinfo{year}{2003}).

\bibitem[{\citenamefont{Maxwell et~al.}(2005)\citenamefont{Maxwell, Brahms,
  deCarvalho, Glenn, Helton, Nguyen, Patterson, Petricka, DeMille, and
  Doyle}}]{Maxwell:PRL95:173201}
\bibinfo{author}{\bibfnamefont{S.~E.} \bibnamefont{Maxwell}},
  \bibinfo{author}{\bibfnamefont{N.}~\bibnamefont{Brahms}},
  \bibinfo{author}{\bibfnamefont{R.}~\bibnamefont{deCarvalho}},
  \bibinfo{author}{\bibfnamefont{D.~R.} \bibnamefont{Glenn}},
  \bibinfo{author}{\bibfnamefont{J.~S.} \bibnamefont{Helton}},
  \bibinfo{author}{\bibfnamefont{S.~V.} \bibnamefont{Nguyen}},
  \bibinfo{author}{\bibfnamefont{D.}~\bibnamefont{Patterson}},
  \bibinfo{author}{\bibfnamefont{J.}~\bibnamefont{Petricka}},
  \bibinfo{author}{\bibfnamefont{D.}~\bibnamefont{DeMille}}, \bibnamefont{and}
  \bibinfo{author}{\bibfnamefont{J.~M.} \bibnamefont{Doyle}},
  \bibinfo{journal}{Phys. Rev. Lett.} \textbf{\bibinfo{volume}{95}},
  \bibinfo{eid}{173201} (\bibinfo{year}{2005}).

\bibitem[{\citenamefont{Volpi and Bohn}(2002)}]{Volpi:PRA65:052712}
\bibinfo{author}{\bibfnamefont{A.}~\bibnamefont{Volpi}} \bibnamefont{and}
  \bibinfo{author}{\bibfnamefont{J.~L.} \bibnamefont{Bohn}},
  \bibinfo{journal}{Physical Review A} \textbf{\bibinfo{volume}{65}},
  \bibinfo{pages}{052712} (\bibinfo{year}{2002}).

\bibitem[{\citenamefont{Krems et~al.}(2003{\natexlab{a}})\citenamefont{Krems,
  Dalgarno, Balakrishnan, and Groenenboom}}]{Krems:PRA67:060703}
\bibinfo{author}{\bibfnamefont{R.~V.} \bibnamefont{Krems}},
  \bibinfo{author}{\bibfnamefont{A.}~\bibnamefont{Dalgarno}},
  \bibinfo{author}{\bibfnamefont{N.}~\bibnamefont{Balakrishnan}},
  \bibnamefont{and} \bibinfo{author}{\bibfnamefont{G.~C.}
  \bibnamefont{Groenenboom}}, \bibinfo{journal}{Phys. Rev. A}
  \textbf{\bibinfo{volume}{67}}, \bibinfo{pages}{060703}
  (\bibinfo{year}{2003}{\natexlab{a}}).

\bibitem[{\citenamefont{Krems et~al.}(2003{\natexlab{b}})\citenamefont{Krems,
  Sadeghpour, Dalgarno, Zgid, Klos, and Chalasinski }}]{Krems:PRA68:051401}
\bibinfo{author}{\bibfnamefont{R.~V.} \bibnamefont{Krems}},
  \bibinfo{author}{\bibfnamefont{H.~R.} \bibnamefont{Sadeghpour}},
  \bibinfo{author}{\bibfnamefont{A.}~\bibnamefont{Dalgarno}},
  \bibinfo{author}{\bibfnamefont{D.}~\bibnamefont{Zgid}},
  \bibinfo{author}{\bibfnamefont{J.}~\bibnamefont{Klos}}, \bibnamefont{and}
  \bibinfo{author}{\bibfnamefont{G.}~\bibnamefont{Chalasinski }},
  \bibinfo{journal}{Phys. Rev. A} \textbf{\bibinfo{volume}{68}},
  \bibinfo{pages}{051401} (\bibinfo{year}{2003}{\natexlab{b}}).

\bibitem[{\citenamefont{Krems and Dalgarno}(2002)}]{Krems:JCP117:118}
\bibinfo{author}{\bibfnamefont{R.}~\bibnamefont{Krems}} \bibnamefont{and}
  \bibinfo{author}{\bibfnamefont{A.}~\bibnamefont{Dalgarno}},
  \bibinfo{journal}{J. Chem. Phys.} \textbf{\bibinfo{volume}{117}},
  \bibinfo{pages}{118} (\bibinfo{year}{2002}).

\bibitem[{\citenamefont{Cybulski et~al.}(2005)\citenamefont{Cybulski, Krems,
  Sadeghpour, Dalgarno, Klos, Groenenboom, van~der Avoird, Zgid, and
  Chalasinski}}]{Cybulski:JCP122:094307}
\bibinfo{author}{\bibfnamefont{H.}~\bibnamefont{Cybulski}},
  \bibinfo{author}{\bibfnamefont{R.~V.} \bibnamefont{Krems}},
  \bibinfo{author}{\bibfnamefont{H.~R.} \bibnamefont{Sadeghpour}},
  \bibinfo{author}{\bibfnamefont{A.}~\bibnamefont{Dalgarno}},
  \bibinfo{author}{\bibfnamefont{J.}~\bibnamefont{Klos}},
  \bibinfo{author}{\bibfnamefont{G.~C.} \bibnamefont{Groenenboom}},
  \bibinfo{author}{\bibfnamefont{A.}~\bibnamefont{van~der Avoird}},
  \bibinfo{author}{\bibfnamefont{D.}~\bibnamefont{Zgid}}, \bibnamefont{and}
  \bibinfo{author}{\bibfnamefont{G.}~\bibnamefont{Chalasinski}},
  \bibinfo{journal}{J. Chem. Phys.} \textbf{\bibinfo{volume}{122}},
  \bibinfo{pages}{094307} (\bibinfo{year}{2005}).

\bibitem[{\citenamefont{Groenenboom}()}]{Groenenboom:privcom}
\bibinfo{author}{\bibfnamefont{G.}~\bibnamefont{Groenenboom}},
  \emph{\bibinfo{title}{private communications}}.

\bibitem[{\citenamefont{Krems and Dalgarno}(2004)}]{Krems:JCP120:2296}
\bibinfo{author}{\bibfnamefont{R.~V.} \bibnamefont{Krems}} \bibnamefont{and}
  \bibinfo{author}{\bibfnamefont{A.}~\bibnamefont{Dalgarno}},
  \bibinfo{journal}{J. Chem. Phys.} \textbf{\bibinfo{volume}{120}},
  \bibinfo{pages}{2296} (\bibinfo{year}{2004}).

\bibitem[{\citenamefont{Maussang et~al.}(2005)\citenamefont{Maussang, Egorov,
  Helton, Nguyen, and Doyle}}]{Maussang:PRL94:123002}
\bibinfo{author}{\bibfnamefont{K.}~\bibnamefont{Maussang}},
  \bibinfo{author}{\bibfnamefont{D.}~\bibnamefont{Egorov}},
  \bibinfo{author}{\bibfnamefont{J.~S.} \bibnamefont{Helton}},
  \bibinfo{author}{\bibfnamefont{S.~V.} \bibnamefont{Nguyen}},
  \bibnamefont{and} \bibinfo{author}{\bibfnamefont{J.~M.} \bibnamefont{Doyle}},
  \bibinfo{journal}{Phys. Rev. Lett.} \textbf{\bibinfo{volume}{94}},
  \bibinfo{pages}{123002} (\bibinfo{year}{2005}).

\bibitem[{\citenamefont{Kaledin et~al.}(1994)\citenamefont{Kaledin, Erickson,
  and Heaven}}]{Kaledin:JMS165:323}
\bibinfo{author}{\bibfnamefont{L.~A.} \bibnamefont{Kaledin}},
  \bibinfo{author}{\bibfnamefont{M.~G.} \bibnamefont{Erickson}},
  \bibnamefont{and} \bibinfo{author}{\bibfnamefont{M.~C.}
  \bibnamefont{Heaven}}, \bibinfo{journal}{J. Mol. Spec.}
  \textbf{\bibinfo{volume}{165}}, \bibinfo{pages}{323} (\bibinfo{year}{1994}).

\bibitem[{\citenamefont{Gordon et~al.}(2005)\citenamefont{Gordon, Appadoo,
  Shayesteh, Walker, and Bernath}}]{Gordon:JMS229:145}
\bibinfo{author}{\bibfnamefont{I.~E.} \bibnamefont{Gordon}},
  \bibinfo{author}{\bibfnamefont{D.~R.~T.} \bibnamefont{Appadoo}},
  \bibinfo{author}{\bibfnamefont{A.}~\bibnamefont{Shayesteh}},
  \bibinfo{author}{\bibfnamefont{K.~A.} \bibnamefont{Walker}},
  \bibnamefont{and} \bibinfo{author}{\bibfnamefont{P.~F.}
  \bibnamefont{Bernath}}, \bibinfo{journal}{J. Mol. Spec.}
  \textbf{\bibinfo{volume}{229}}, \bibinfo{pages}{145} (\bibinfo{year}{2005}).

\bibitem[{\citenamefont{Sheridan and Ziurys}(2003)}]{Sheridan:CPL380:632}
\bibinfo{author}{\bibfnamefont{P.~M.} \bibnamefont{Sheridan}} \bibnamefont{and}
  \bibinfo{author}{\bibfnamefont{L.~M.} \bibnamefont{Ziurys}},
  \bibinfo{journal}{Chem. Phys. Lett.} \textbf{\bibinfo{volume}{380}},
  \bibinfo{pages}{632} (\bibinfo{year}{2003}).

\bibitem[{\citenamefont{Halfen and Ziurys}(2005)}]{Halfen:JCP122:054309}
\bibinfo{author}{\bibfnamefont{D.~T.} \bibnamefont{Halfen}} \bibnamefont{and}
  \bibinfo{author}{\bibfnamefont{L.~M.} \bibnamefont{Ziurys}},
  \bibinfo{journal}{J. Chem. Phys.} \textbf{\bibinfo{volume}{122}},
  \bibinfo{pages}{054309} (\bibinfo{year}{2005}).

\bibitem[{\citenamefont{Corkery et~al.}(1991)\citenamefont{Corkery, Brown,
  Beaton, and Evenson}}]{Corkery:JMS149:257}
\bibinfo{author}{\bibfnamefont{S.~M.} \bibnamefont{Corkery}},
  \bibinfo{author}{\bibfnamefont{J.~M.} \bibnamefont{Brown}},
  \bibinfo{author}{\bibfnamefont{S.~P.} \bibnamefont{Beaton}},
  \bibnamefont{and} \bibinfo{author}{\bibfnamefont{K.~M.}
  \bibnamefont{Evenson}}, \bibinfo{journal}{J. Mol. Spec.}
  \textbf{\bibinfo{volume}{149}}, \bibinfo{pages}{257} (\bibinfo{year}{1991}).

\bibitem[{\citenamefont{Namiki and Saito}(1997)}]{Namiki:JCP107:8848}
\bibinfo{author}{\bibfnamefont{K.}~\bibnamefont{Namiki}} \bibnamefont{and}
  \bibinfo{author}{\bibfnamefont{S.}~\bibnamefont{Saito}}, \bibinfo{journal}{J.
  Chem. Phys.} \textbf{\bibinfo{volume}{107}}, \bibinfo{pages}{8848}
  (\bibinfo{year}{1997}).

\bibitem[{\citenamefont{Katoh et~al.}(2004)\citenamefont{Katoh, Okabayashi,
  Tanimoto, Sumiyoshi, and Endo}}]{Katoh:JCP120:7927}
\bibinfo{author}{\bibfnamefont{K.}~\bibnamefont{Katoh}},
  \bibinfo{author}{\bibfnamefont{T.}~\bibnamefont{Okabayashi}},
  \bibinfo{author}{\bibfnamefont{M.}~\bibnamefont{Tanimoto}},
  \bibinfo{author}{\bibfnamefont{Y.}~\bibnamefont{Sumiyoshi}},
  \bibnamefont{and} \bibinfo{author}{\bibfnamefont{Y.}~\bibnamefont{Endo}},
  \bibinfo{journal}{J. Chem. Phys.} \textbf{\bibinfo{volume}{120}},
  \bibinfo{pages}{7927} (\bibinfo{year}{2004}).

\bibitem[{\citenamefont{Thompsen et~al.}(2002)\citenamefont{Thompsen, Brewster,
  and Ziurys}}]{Thompsen:JCP116:10212}
\bibinfo{author}{\bibfnamefont{J.~M.} \bibnamefont{Thompsen}},
  \bibinfo{author}{\bibfnamefont{M.~A.} \bibnamefont{Brewster}},
  \bibnamefont{and} \bibinfo{author}{\bibfnamefont{L.~M.}
  \bibnamefont{Ziurys}}, \bibinfo{journal}{J. Chem. Phys.}
  \textbf{\bibinfo{volume}{116}}, \bibinfo{pages}{10212}
  (\bibinfo{year}{2002}).

\bibitem[{\citenamefont{Sheridan et~al.}(2002)\citenamefont{Sheridan, Brewster,
  and Ziurys}}]{Sheridan:ApJ576:1108}
\bibinfo{author}{\bibfnamefont{P.~M.} \bibnamefont{Sheridan}},
  \bibinfo{author}{\bibfnamefont{M.~A.} \bibnamefont{Brewster}},
  \bibnamefont{and} \bibinfo{author}{\bibfnamefont{L.~M.}
  \bibnamefont{Ziurys}}, \bibinfo{journal}{Astrophys. J.}
  \textbf{\bibinfo{volume}{576}}, \bibinfo{pages}{1108} (\bibinfo{year}{2002}).

\bibitem[{\citenamefont{Cheung et~al.}(1982)\citenamefont{Cheung, Hansen, and
  Merer}}]{Cheung:JMS91:165}
\bibinfo{author}{\bibfnamefont{A.~S.~C.} \bibnamefont{Cheung}},
  \bibinfo{author}{\bibfnamefont{R.~C.} \bibnamefont{Hansen}},
  \bibnamefont{and} \bibinfo{author}{\bibfnamefont{A.~J.} \bibnamefont{Merer}},
  \bibinfo{journal}{J. Mol. Spec.} \textbf{\bibinfo{volume}{91}},
  \bibinfo{pages}{165} (\bibinfo{year}{1982}).

\bibitem[{\citenamefont{Adam et~al.}(1994)\citenamefont{Adam, Azuma, Barry,
  Merer, Sassenberg, Schroder, Cheval, and Femenias}}]{Adam:JCP100:6240}
\bibinfo{author}{\bibfnamefont{A.~G.} \bibnamefont{Adam}},
  \bibinfo{author}{\bibfnamefont{Y.}~\bibnamefont{Azuma}},
  \bibinfo{author}{\bibfnamefont{J.~A.} \bibnamefont{Barry}},
  \bibinfo{author}{\bibfnamefont{A.~J.} \bibnamefont{Merer}},
  \bibinfo{author}{\bibfnamefont{U.}~\bibnamefont{Sassenberg}},
  \bibinfo{author}{\bibfnamefont{J.~O.} \bibnamefont{Schroder}},
  \bibinfo{author}{\bibfnamefont{G.}~\bibnamefont{Cheval}}, \bibnamefont{and}
  \bibinfo{author}{\bibfnamefont{J.~L.} \bibnamefont{Femenias}},
  \bibinfo{journal}{J. Chem. Phys.} \textbf{\bibinfo{volume}{100}},
  \bibinfo{pages}{6240} (\bibinfo{year}{1994}).

\bibitem[{\citenamefont{Launila and Lindgren}(1996)}]{Launila:JCP104:6418}
\bibinfo{author}{\bibfnamefont{O.}~\bibnamefont{Launila}} \bibnamefont{and}
  \bibinfo{author}{\bibfnamefont{B.}~\bibnamefont{Lindgren}},
  \bibinfo{journal}{J. Chem. Phys.} \textbf{\bibinfo{volume}{104}},
  \bibinfo{pages}{6418} (\bibinfo{year}{1996}).

\bibitem[{\citenamefont{Brazier et~al.}(1986)\citenamefont{Brazier, Ram, and
  Bernath}}]{Brazier:jms120:381}
\bibinfo{author}{\bibfnamefont{C.~R.} \bibnamefont{Brazier}},
  \bibinfo{author}{\bibfnamefont{R.~S.} \bibnamefont{Ram}}, \bibnamefont{and}
  \bibinfo{author}{\bibfnamefont{P.~F.} \bibnamefont{Bernath}},
  \bibinfo{journal}{J. Mol. Spec.} \textbf{\bibinfo{volume}{120}},
  \bibinfo{pages}{381} (\bibinfo{year}{1986}).

\bibitem[{\citenamefont{Berden et~al.}(1998)\citenamefont{Berden, Engeln,
  Christianen, Maan, and Meijer}}]{Berden:PRA58:3114}
\bibinfo{author}{\bibfnamefont{G.}~\bibnamefont{Berden}},
  \bibinfo{author}{\bibfnamefont{R.}~\bibnamefont{Engeln}},
  \bibinfo{author}{\bibfnamefont{P.~C.~M.} \bibnamefont{Christianen}},
  \bibinfo{author}{\bibfnamefont{J.~C.} \bibnamefont{Maan}}, \bibnamefont{and}
  \bibinfo{author}{\bibfnamefont{G.}~\bibnamefont{Meijer}},
  \bibinfo{journal}{Phys. Rev. A} \textbf{\bibinfo{volume}{58}},
  \bibinfo{pages}{3114} (\bibinfo{year}{1998}).

\bibitem[{\citenamefont{Barclay et~al.}(1993)\citenamefont{Barclay, Anderson,
  and Ziurys}}]{Barclay:ApJ408:L65}
\bibinfo{author}{\bibfnamefont{W.~L.} \bibnamefont{Barclay}},
  \bibinfo{author}{\bibfnamefont{M.~A.} \bibnamefont{Anderson}},
  \bibnamefont{and} \bibinfo{author}{\bibfnamefont{L.~M.}
  \bibnamefont{Ziurys}}, \bibinfo{journal}{Astrophys. J.}
  \textbf{\bibinfo{volume}{408}}, \bibinfo{pages}{L65} (\bibinfo{year}{1993}).

\bibitem[{\citenamefont{Gaydon and Pearse}(1937)}]{Gaydon:Nat140:110}
\bibinfo{author}{\bibfnamefont{A.~G.} \bibnamefont{Gaydon}} \bibnamefont{and}
  \bibinfo{author}{\bibfnamefont{R.~W.~B.} \bibnamefont{Pearse}},
  \bibinfo{journal}{Nature} \textbf{\bibinfo{volume}{140}},
  \bibinfo{pages}{110} (\bibinfo{year}{1937}).

\bibitem[{\citenamefont{Pearse and Gaydon}(1938)}]{Pearse:ProcPhysSoc50:201}
\bibinfo{author}{\bibfnamefont{R.~W.~B.} \bibnamefont{Pearse}}
  \bibnamefont{and} \bibinfo{author}{\bibfnamefont{A.~G.}
  \bibnamefont{Gaydon}}, \bibinfo{journal}{Proc. Phys. Soc.}
  \textbf{\bibinfo{volume}{50}}, \bibinfo{pages}{201} (\bibinfo{year}{1938}).

\bibitem[{\citenamefont{Kirkpatrick et~al.}(1999)\citenamefont{Kirkpatrick,
  Reid, Liebert, Cutri, Nelson, Beichman, Dahn, Monet, Gizis, and
  Skrutskie}}]{Kirkpatrick:APJ519:802}
\bibinfo{author}{\bibfnamefont{J.~D.} \bibnamefont{Kirkpatrick}},
  \bibinfo{author}{\bibfnamefont{I.~N.} \bibnamefont{Reid}},
  \bibinfo{author}{\bibfnamefont{J.}~\bibnamefont{Liebert}},
  \bibinfo{author}{\bibfnamefont{R.~M.} \bibnamefont{Cutri}},
  \bibinfo{author}{\bibfnamefont{B.}~\bibnamefont{Nelson}},
  \bibinfo{author}{\bibfnamefont{C.~A.} \bibnamefont{Beichman}},
  \bibinfo{author}{\bibfnamefont{C.~C.} \bibnamefont{Dahn}},
  \bibinfo{author}{\bibfnamefont{D.~G.} \bibnamefont{Monet}},
  \bibinfo{author}{\bibfnamefont{J.~E.} \bibnamefont{Gizis}}, \bibnamefont{and}
  \bibinfo{author}{\bibfnamefont{M.~F.} \bibnamefont{Skrutskie}},
  \bibinfo{journal}{Astrophys. J.} \textbf{\bibinfo{volume}{519}},
  \bibinfo{pages}{802} (\bibinfo{year}{1999}).

\bibitem[{\citenamefont{Shin et~al.}(2005)\citenamefont{Shin, Brugh, and
  Morse}}]{Shin:ApJ619:407}
\bibinfo{author}{\bibfnamefont{S.}~\bibnamefont{Shin}},
  \bibinfo{author}{\bibfnamefont{D.~J.} \bibnamefont{Brugh}}, \bibnamefont{and}
  \bibinfo{author}{\bibfnamefont{M.~D.} \bibnamefont{Morse}},
  \bibinfo{journal}{Astrophys. J.} \textbf{\bibinfo{volume}{619}},
  \bibinfo{pages}{407} (\bibinfo{year}{2005}).

\bibitem[{\citenamefont{Engvold et~al.}(1980)\citenamefont{Engvold, W{\"o}hl,
  and Brault}}]{Engvold:AAsuplser42:209}
\bibinfo{author}{\bibfnamefont{O.}~\bibnamefont{Engvold}},
  \bibinfo{author}{\bibfnamefont{H.}~\bibnamefont{W{\"o}hl}}, \bibnamefont{and}
  \bibinfo{author}{\bibfnamefont{J.}~\bibnamefont{Brault}},
  \bibinfo{journal}{Astron.Astrophys.Suppl.Series}
  \textbf{\bibinfo{volume}{42}}, \bibinfo{pages}{209} (\bibinfo{year}{1980}).

\bibitem[{\citenamefont{Chen et~al.}()\citenamefont{Chen, Bakker, Peters,
  Stoll, Meijer, and Steimle}}]{Chen-inprep}
\bibinfo{author}{\bibfnamefont{J.}~\bibnamefont{Chen}},
  \bibinfo{author}{\bibfnamefont{J.~M.} \bibnamefont{Bakker}},
  \bibinfo{author}{\bibfnamefont{A.}~\bibnamefont{Peters}},
  \bibinfo{author}{\bibfnamefont{M.}~\bibnamefont{Stoll}},
  \bibinfo{author}{\bibfnamefont{G.}~\bibnamefont{Meijer}}, \bibnamefont{and}
  \bibinfo{author}{\bibfnamefont{T.~C.} \bibnamefont{Steimle}},
  \bibinfo{journal}{to be published}  (????).

\bibitem[{\citenamefont{Varberg et~al.}(1990)\citenamefont{Varberg, Field, and
  Merer}}]{Varberg:jcp92:7123}
\bibinfo{author}{\bibfnamefont{T.~D.} \bibnamefont{Varberg}},
  \bibinfo{author}{\bibfnamefont{R.~W.} \bibnamefont{Field}}, \bibnamefont{and}
  \bibinfo{author}{\bibfnamefont{A.~J.} \bibnamefont{Merer}},
  \bibinfo{journal}{J. Chem. Phys.} \textbf{\bibinfo{volume}{92}},
  \bibinfo{pages}{7123} (\bibinfo{year}{1990}).

\bibitem[{\citenamefont{Varberg et~al.}(1991)\citenamefont{Varberg, Field, and
  Merer}}]{Varberg:JCP95:1563}
\bibinfo{author}{\bibfnamefont{T.~D.} \bibnamefont{Varberg}},
  \bibinfo{author}{\bibfnamefont{R.~W.} \bibnamefont{Field}}, \bibnamefont{and}
  \bibinfo{author}{\bibfnamefont{A.~J.} \bibnamefont{Merer}},
  \bibinfo{journal}{J. Chem. Phys.} \textbf{\bibinfo{volume}{95}},
  \bibinfo{pages}{1563} (\bibinfo{year}{1991}).

\bibitem[{\citenamefont{Varberg et~al.}(1992)\citenamefont{Varberg, Gray,
  Field, and Merer}}]{Varberg:JMS156:296}
\bibinfo{author}{\bibfnamefont{T.~D.} \bibnamefont{Varberg}},
  \bibinfo{author}{\bibfnamefont{J.~A.} \bibnamefont{Gray}},
  \bibinfo{author}{\bibfnamefont{R.~W.} \bibnamefont{Field}}, \bibnamefont{and}
  \bibinfo{author}{\bibfnamefont{A.~J.} \bibnamefont{Merer}},
  \bibinfo{journal}{J. Mol. Spec.} \textbf{\bibinfo{volume}{156}},
  \bibinfo{pages}{296} (\bibinfo{year}{1992}).

\bibitem[{\citenamefont{Vanzee et~al.}(1986{\natexlab{a}})\citenamefont{Vanzee,
  Garland, and Weltner}}]{Vanzee:JCP84:5968}
\bibinfo{author}{\bibfnamefont{R.~J.} \bibnamefont{Vanzee}},
  \bibinfo{author}{\bibfnamefont{D.~A.} \bibnamefont{Garland}},
  \bibnamefont{and} \bibinfo{author}{\bibfnamefont{W.}~\bibnamefont{Weltner}},
  \bibinfo{journal}{J. Chem. Phys.} \textbf{\bibinfo{volume}{84}},
  \bibinfo{pages}{5968} (\bibinfo{year}{1986}{\natexlab{a}}).

\bibitem[{\citenamefont{Vanzee et~al.}(1986{\natexlab{b}})\citenamefont{Vanzee,
  Garland, and Weltner}}]{Vanzee:JCP85:3237}
\bibinfo{author}{\bibfnamefont{R.~J.} \bibnamefont{Vanzee}},
  \bibinfo{author}{\bibfnamefont{D.~A.} \bibnamefont{Garland}},
  \bibnamefont{and} \bibinfo{author}{\bibfnamefont{W.}~\bibnamefont{Weltner}},
  \bibinfo{journal}{J. Chem. Phys.} \textbf{\bibinfo{volume}{85}},
  \bibinfo{pages}{3237} (\bibinfo{year}{1986}{\natexlab{b}}).

\end{thebibliography}

\newpage

\end{document}